\documentclass[12pt]{article}
\usepackage{geometry,hyperref} % see geometry.pdf on how to lay out the page. There's lots.
\usepackage{graphicx}				% Use pdf, png, jpg, or eps§ with pdflatex; use eps in DVI mode
\usepackage{amssymb}
\usepackage{lmodern}
\usepackage[symbol]{footmisc}
\newcommand*{\dprime}{^{\prime\prime}\mkern-1.2mu}

\begin{document}
% See the ``Article customise'' template for come common customisations

\begin{center}

{\Large My companion is bigger than your companion!}

\vspace{0.5cm}

\emph{
By Henri M.J. Boffin\\European Southern Observatory, Karl-Schwarzschild-str. 2, 85748 Garching, Germany,\\
and\\
Virginia Trimble\\University of California, Irvine, CA 92697-4575, and Las Cumbres Observatory
}

\end {center}

%%% BEGIN DOCUMENT

{\bf We provide an analysis of the mass ratio distribution as gathered from almost all of the 559 orbital solutions derived by Professor Roger Griffin in his long series in the \emph{Observatory Magazine} about `Spectroscopic Binary Orbits from Photoelectric Radial Velocities'. The total distribution we determine is close to a uniform one, with a dearth of the smallest companions and an excess of almost twins. When splitting our sample between main-sequence and red giant primaries, however, we discover a different picture: the excess of twins is limited to the main-sequence stars, for which it appears even more pronounced. The mass-ratio distributions of red giants is characterised by a decline of systems with mass ratio above 0.6 and an excess of systems with a mass ratio around 0.25, which we attribute to post-mass transfer systems. The difference between the two mass-ratio distributions is likely due to the different primary masses they sample.}

\vspace{0.5cm}
\emph{A brief outline of a long career}
\\

%{\it Here we could consider putting an introductory section that covers the items mentioned by the section subtitle, but also about Roger Griffin, his world record series in the Magazine, as well as the importance of the mass ratio distribution.}
%\\

Cambridge Professor Emeritus Roger Francis Griffin, whose extraordinary 45 years of studying spectroscopic binary stars form the basis for this analysis, is a native of Surrey, England. His early education came from the Caterham School (plus a certain amount of ``home schooling'', because his mother had been a teacher). Roger received his PhD in 1960 from Cambridge University for work with Prof. Roderick Olivier Redman (first director of the combined Observatories of Cambridge) on narrow-band stellar photometry, which was never fully published, using a narrow-band spectrometer of their own devising. A visiting scholarship (we would probably now say postdoctoral fellowship) at the Radcliffe Observatory in Pretoria, South Africa, with A. David Thackeray yielded spectroscopy of Nova RS Oph with photographic plates, which Griffin already knew were not the most efficient technology for the purpose. A second excursion from 1961 to Pasadena (Caltech and Mt. Wilson Observatory) was supported by a Carnegie Fellowship for a project that turned out not amenable to research. He participated in optical identification of 3C radio sources using plates from the 1.2-m (48$\dprime$) Schmidt and 5-m (200$\dprime$) Hale telescopes, then turned to the 2.5-m (100$\dprime$) Hooker telescope to begin taking extraordinarily high-dispersion spectra of Arcturus. He returned to the 2.5-m to finish this after being appointed a Fellow of St. John's College. 
The tracings went to a new graduate student at Cambridge, Rita Elizabeth Mary Gasson in 1963 for her PhD work, which resulted in the Arcturus Atlas$^1$, her PhD degree in 1966, and joint papers from the next year onward. They married in 1966 in Sussex, had two sons, Rupert and Richard, and divorced in 2002\footnote{In the honeymoon years, she was Cindy, he was Yogi, and the Cambridge home they shared was called Jellystone.}. The  Arcturus Atlas is Roger's second-most cited paper. Their joint work on the Procyon Atlas$^2$ recognised the value of using telluric lines for precise radial velocity calibration$^3$.

Roger's explanation$^4$ of how to do a better job of measuring radial velocities using a photomultiplier tube and coded mask as a radial velocity spectrometer\footnote{His friends secretly call these {\it Griffinometers}.} remains, at 290, his most cited paper. It includes a brief but inclusive history of stellar radial velocity measurements in general and the method in particular 
(with due credit to Babcock and Fellgett, who, however, never 
built
the relevant widgets).  Babcock was, however, the director at Mt. Wilson-Palomar Observatories when
Roger F. Griffin and James E. Gunn built a radial velocity spectrometer (including 10 Lego motors) 
to operate on the Palomar 5-m and study stars in globular clusters
(no binaries to speak of, and no dark matter) and such.  Paper 1 (Arabic, not Roman numerals, though he had already
used Roman numerals for another series of papers) appeared in {\it The Observatory}$^5$
in 1975 with co-author B. Emerson (of the Royal Greenwich Observatory).  A successor appeared in
every issue of {\it The Observatory} until recently, a few with co-authors
(including the present first author), but most as single-author papers,
written in a unique style that you must experience for yourself.  He
is the only editor ever to have modified a paper by the present second
author without arousing her eternal enmity.
The series has now reached paper 265,
and we thought it time to take another look at the distribution of
binary system mass ratios in this remarkable sample.  %A previous effort by the junior author (ref) annoyed the senior author sufficiently that working together seemed the obvious solution. %
We do this with the
gracious permission of Prof. Roger Griffin and some biographical assistance
from Dr. Elizabeth Griffin.  They both objected to the loss of the address
The Observatories, Cambridge, when it merged with Hoyle's Institute of Theoretical
Astronomy to become the Institute of Astronomy.

Some of the observing was done with the Coravel at the Haute-Provence Observatory, and some of the devices with which extrasolar planets have been discovered by radial velocity measurements make use of both the photoelectric photometer with coded-mask concept and the use of telluric lines for wavelength calibration.

%\vspace{1cm}
\pagebreak
\emph{A revised look at the mass-ratio distribution}
\\
More  than a decade ago, on the occasion of the publication of the 200$^{th}$ paper in Griffin's Series, one of us presented an analysis$^6$ of the mass ratios that derived from these data, supplemented from additional orbits published by Griffin in other publications. As we have now reached number 265 in the series -- a number we hope will continue to grow, the most critical reader may wonder why is there a need for another such analysis \emph{now}. One reason might be that the methodology used by the present second
author of the current paper was questioned by, among others, the
current first author$^7$, and working together seemed the obvious solution. Another reason, and more important, is that the last decade has seen in the papers of the \emph{Series} a change in ground rules from a strict one star
per paper to, often, as many as 4--6 per paper and in one case, even 10. Thus, if in the Papers 1--200 of the \emph{Series}, there were 290 orbits presented, in the remaining Papers 201--265, there were no fewer than 269! This almost doubling of the number of systems justifies on its own a reanalysis of the sample. Finally, one should also note that since the last analysis, we have now entered the \emph{Gaia} era of Galactic astronomy, that has completely revolutionised our field. 

\begin{figure}[htbp]
   \centering
   \includegraphics[width=1.0\textwidth]{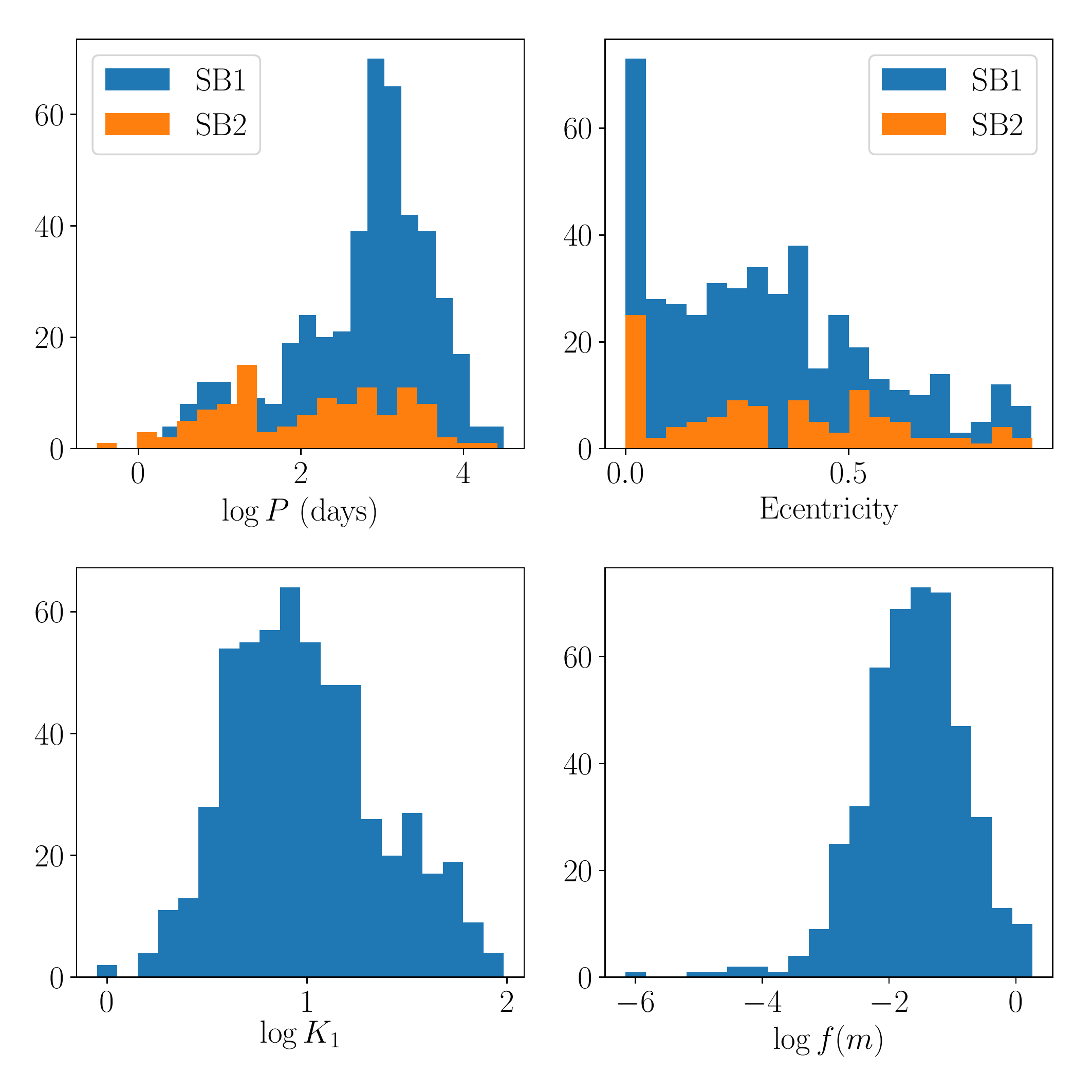} % requires the graphicx package
   \caption{The distribution of the orbital elements for the sample considered.}
   \label{fig:dist}
\end{figure}

\vspace{0.5cm}
\emph{The sample}
\\
The reanalysis is even more warranted as the 559 orbits covered in this Series encompass a most amazing range of the various orbital elements: the orbital period, $P$, ranges from 0.31020614 to 31,292 days\footnote{Not counting one system whose orbital period was fixed at 50,000 days.} ($\sim 86$ years, i.e. more than Griffin's age!), while the non-zero  significant eccentricities, $e$, vary between 0.0078\footnote{Such tiny, but possibly significant eccentricities are often due to the presence of a third component that perturbs the orbit, and indeed in those cases, Griffin either computed an outer orbit or showed that the center of mass velocity of the binary was changing with time -- a clear sign of a third companion.} and 0.9119. Those are by no means record holders, but are quite remarkable indeed and bear witness of Griffin's method to derive orbits: if one star happens to catch his attention for some reason or another, it will be observed on several occasions, over many years, so that if it happens to be a spectroscopic binary, it will be sooner or later unmasked! This is also visible through the range of semi-amplitudes of the primary radial velocity curve that is covered: between 0.89 and 96.27 km/s. The distribution of those elements, shown in Fig.~\ref{fig:dist}, bear a strong resemblance to those shown by Griffin in his own summary$^8$ more than a decade ago. 

The distribution of elements described above leads to a range of spectroscopic mass functions, $f(m)$, for single-lined binaries that is comprised between 0.00000070 (!) and 1.823 M$_\odot$. %The former corresponds to 0.23 Earth masses! 
The methodology used by Griffin and the long time spans that cover these orbits imply that we shouldn't expect any bias in terms of the inclinations of the orbits -- something that will be useful when determining the mass ratios of SB1 systems.

\begin{figure}[htbp]
   \centering
   \includegraphics[width=1.0\textwidth]{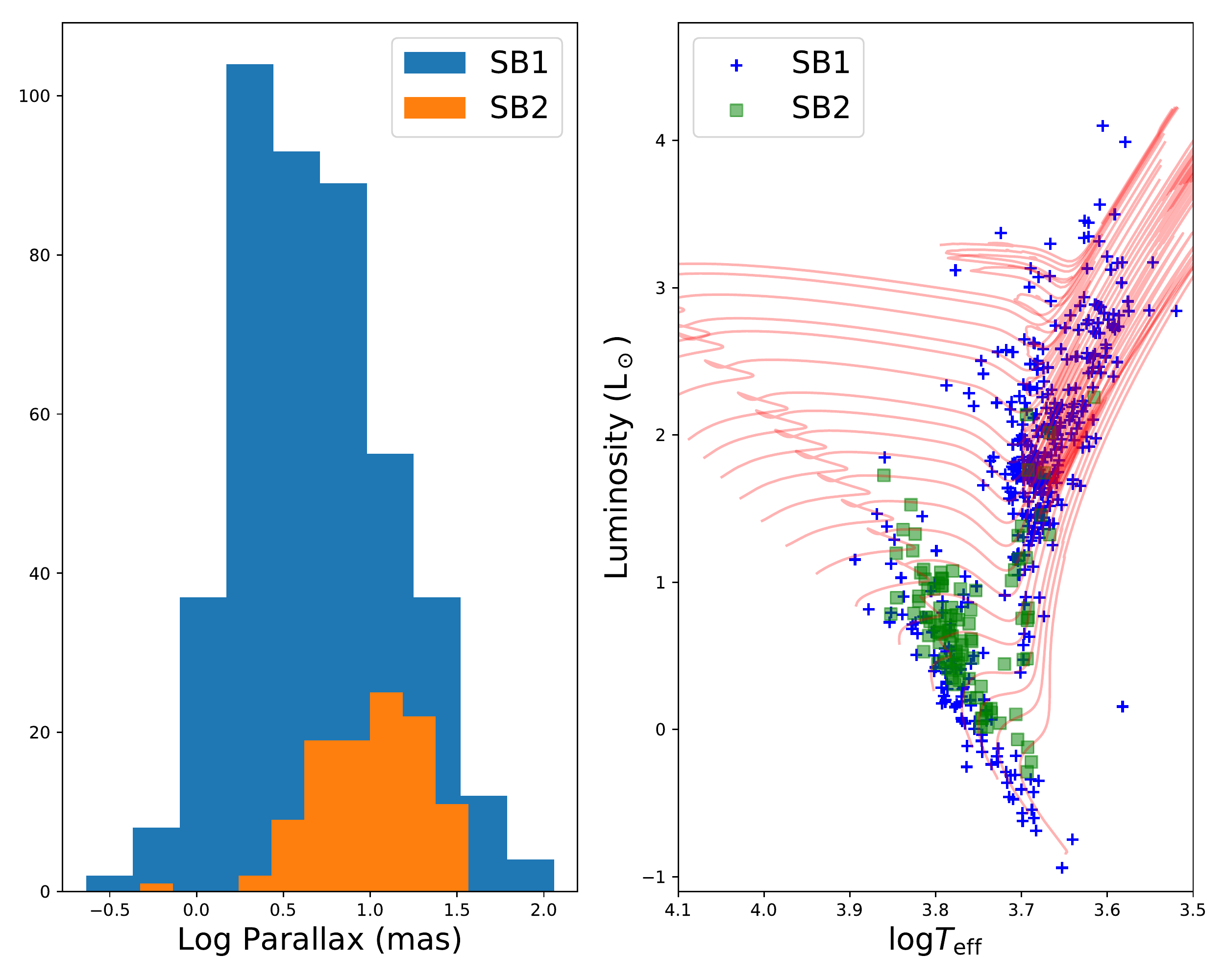} % requires the graphicx package
   \caption{(Left) The distribution in parallaxes of the stars in the initial sample, split among SB1 and SB2 systems. (Right) H-R diagramme of the systems from our final sample based on \emph{Gaia} DR2 effective temperatures and radii. These are compared to MESA stellar evolutionary tracks for a wide range of masses and of solar metallicity. }
   \label{fig:CMD}
\end{figure}

The  sample we extracted from Griffin's papers is comprised of 
111 double-lined spectroscopic binaries (SB2) and 
448 single-lined spectroscopic binaries (SB1).
We have cross-matched this sample with the \emph{Gaia} DR2 data set$^{9,10}$. For most of our targets (431 SB1 and 109 SB2 -- our final sample), the \emph{Gaia} DR2 provides us with the effective temperature and radius, allowing us to place them in an H-R diagramme (Fig.~\ref{fig:CMD}).
From this, it is clear that there aren't many SB2s with a giant component, and that the contribution from the companion in SB2s shift them up and to the red with respect to the tracks.

\vspace{0.5cm}
\emph{Mass ratios of SB2 systems}
\\

For SB2s, deriving the mass ratio is a trivial thing as this is simply the ratio between the two semi-amplitudes of the orbits ($K_1, K_2$) of the components in the system. When doing so in our sample, we note that we have a few systems that have \emph{formally} a mass ratio, $q$, above one. This is at first hand unexpected as the primary is by most definitions the brightest, which in normal stellar evolution, also implies the most massive, unless there has been mass transfer. Although this is a possibility, we note that in all cases where $q > 1$, the error bars make it easily move on the \emph{correct} side of the inequality, within one standard deviation.  We should thus consider that these systems are in fact compatible with being twins ($q \approx 1$). 
It is also useful to note that by the nature of the observations done -- cross-correlation with a mask of the red giant Arcturus -- the primary in Griffin's papers will be the one that gives the deeper dip in his tracings, which is usually the brighter, although for pairs of quite different spectral types, it could be the less luminous but the redder one. A look at 
the spectral types distribution for these systems reveal that most are of F-type.

%{\bf TO DO: Do we see that they are confined to systems with $P < 50$ days?} -- they are not !!!

\begin{figure}[htbp]
   \centering
   \includegraphics[width=0.75\textwidth]{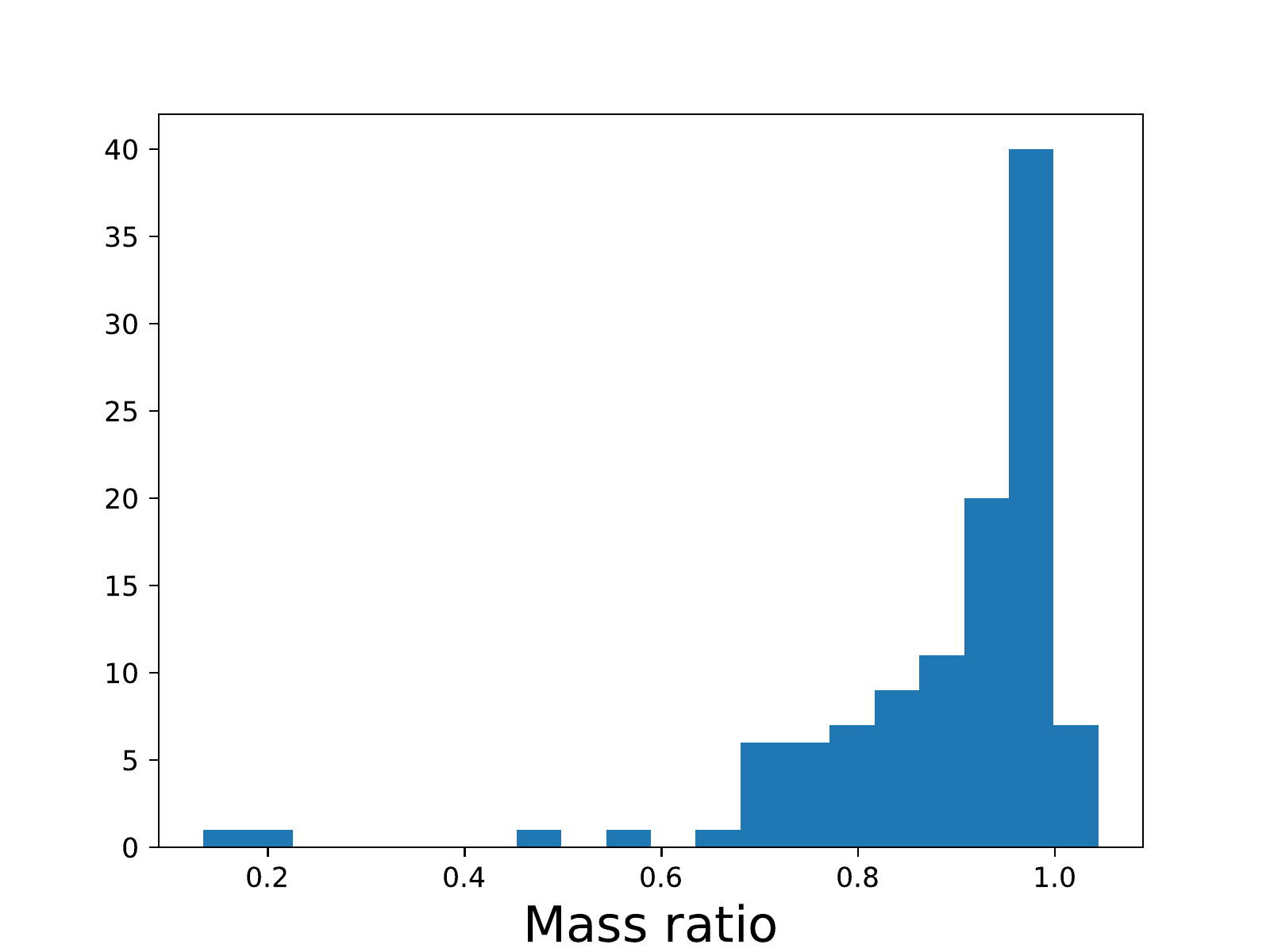} % requires the graphicx package
   \caption{The distribution of the mass ratio for all SB2 systems.}
   \label{fig:sb2q}
\end{figure}

Most of the SB2 have mass ratios between 0.6 and 1. This is expected, as most are on the main-sequence: a system will appear as a SB2 if the difference in brightness between the two is less than about 2--2.5 magnitudes (depending on the colour and the matching with the spectral type of the Arcturus mask), i.e. roughly a factor ten in luminosity (with a clear colour dependence). Given the mass-luminosity relation on the main-sequence for solar-like stars, $L \propto M^{4.5}$, this leads to a minimum mass ratio of about 0.6, as observed. There are, however, four systems with mass ratios below this value:
\begin{itemize}
    \item HD 192785 (q = 0.13; P = 19.2735 d) is clearly an example of a mass transfer system;
    \item HD 31738  (q = 0.21;  P = 0.45 d) was also reported by Griffin has having a ``period [that] is unexpectedly, indeed astoundingly, short," and ``one in which there has been a lot of mass exchange;"
    \item HD 158209 (q = 0.46) is a triple system, where the primary is a binary system containing a subgiant, while the secondary is a main sequence star;
    \item HD 100125 (q = 0.56; P = 48 d) shows a secondary peak so weak that only the experienced eye of Griffin would have seen it, and the secondary orbit is based mostly on data obtained by others at Kitt Peak.  With q=0.56, it is indeed at the limit of what we expect to be an  SB2.
\end{itemize}

\begin{figure}[htbp]
   \centering
   \includegraphics[width=0.75\textwidth]{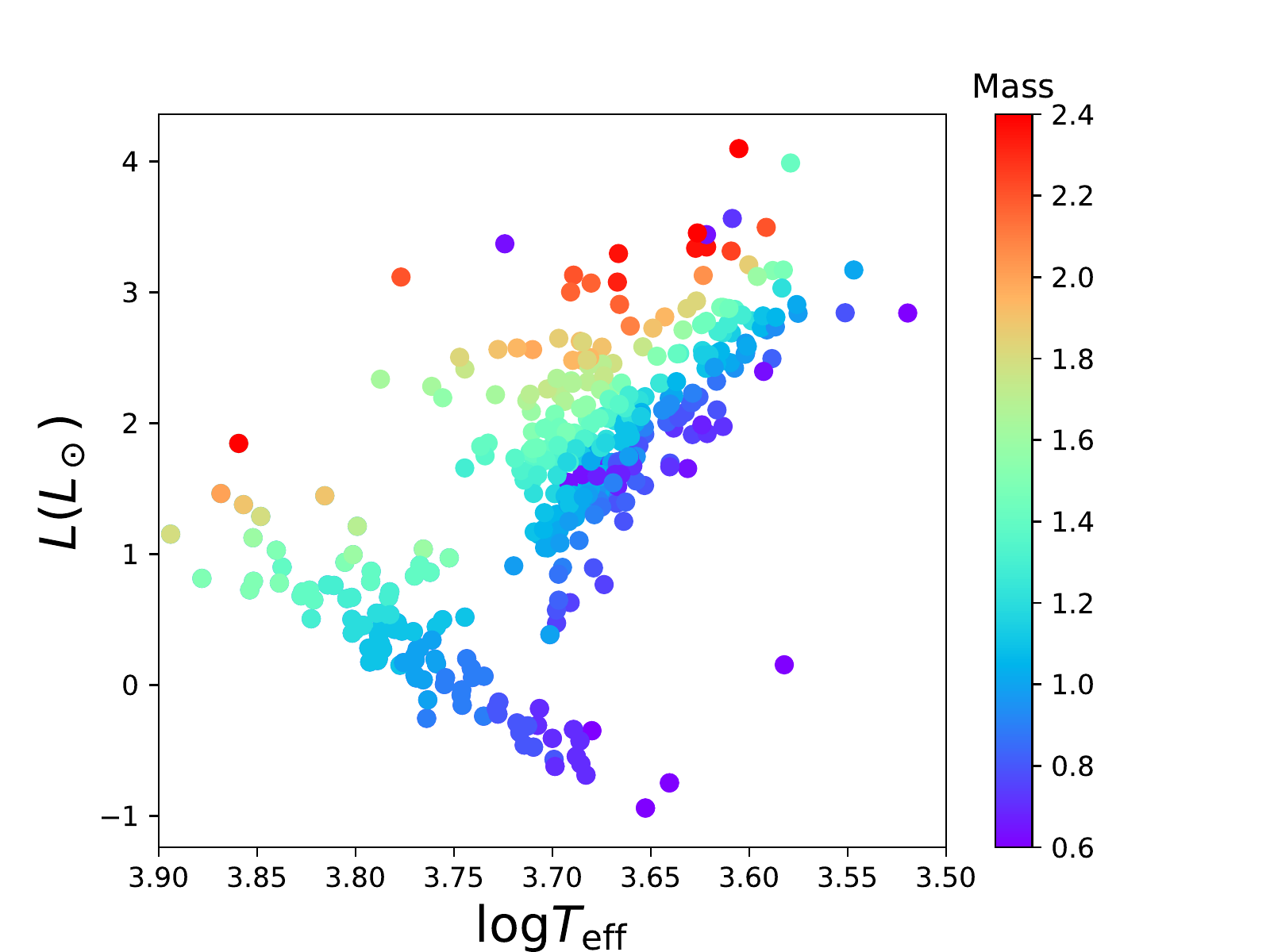} % requires the graphicx package
   \caption{H-R diagram showing all SB1 systems for which \emph{Gaia} DR2 provides temperature and radius. They are greyed based on the primary mass we derived based on MESA evolutionary tracks.}
   \label{fig:hr}
\end{figure}

\begin{figure}[htbp]
   \centering
   \includegraphics[width=0.8\textwidth]{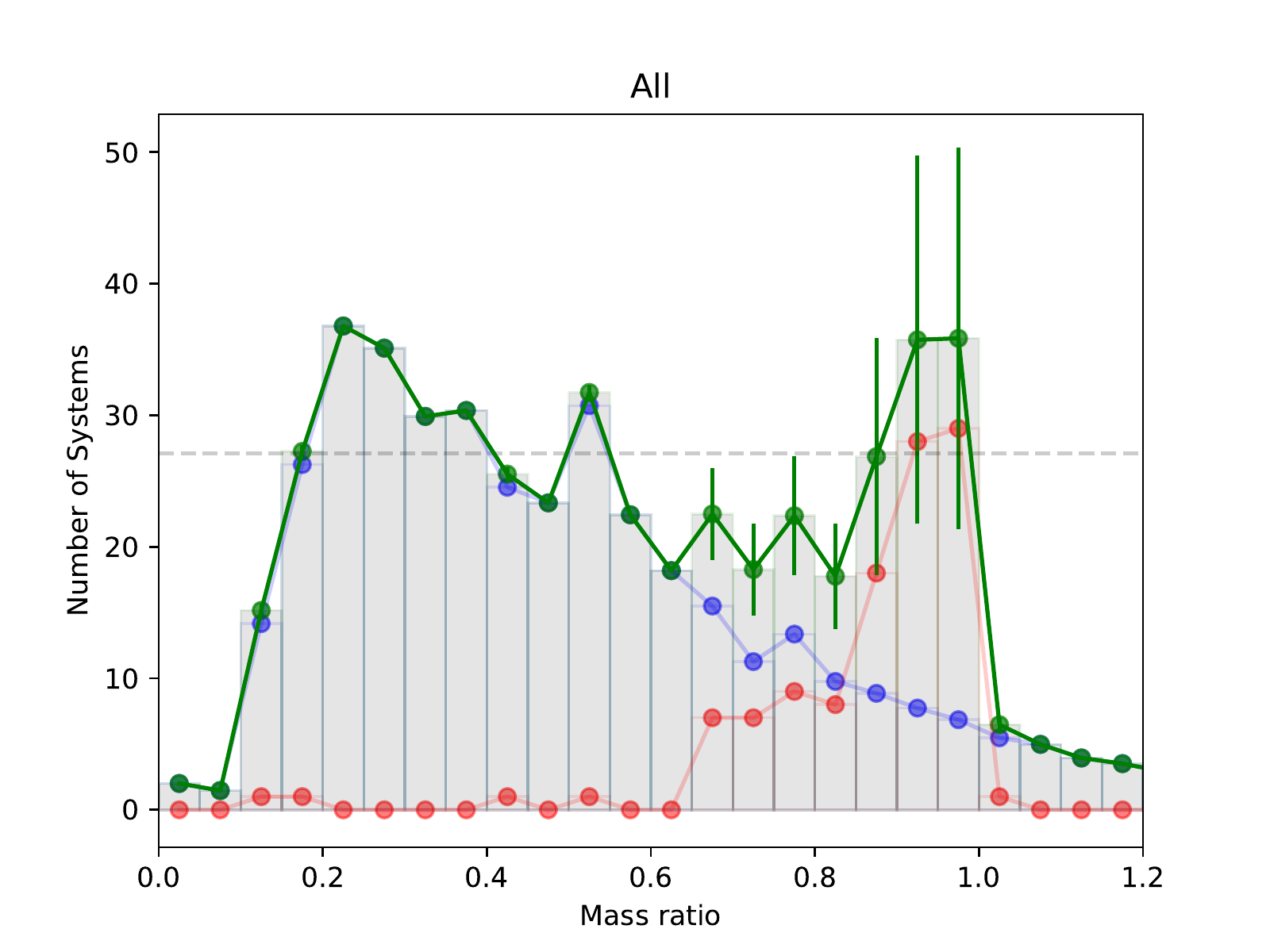}\\
    \includegraphics[width=0.8\textwidth]{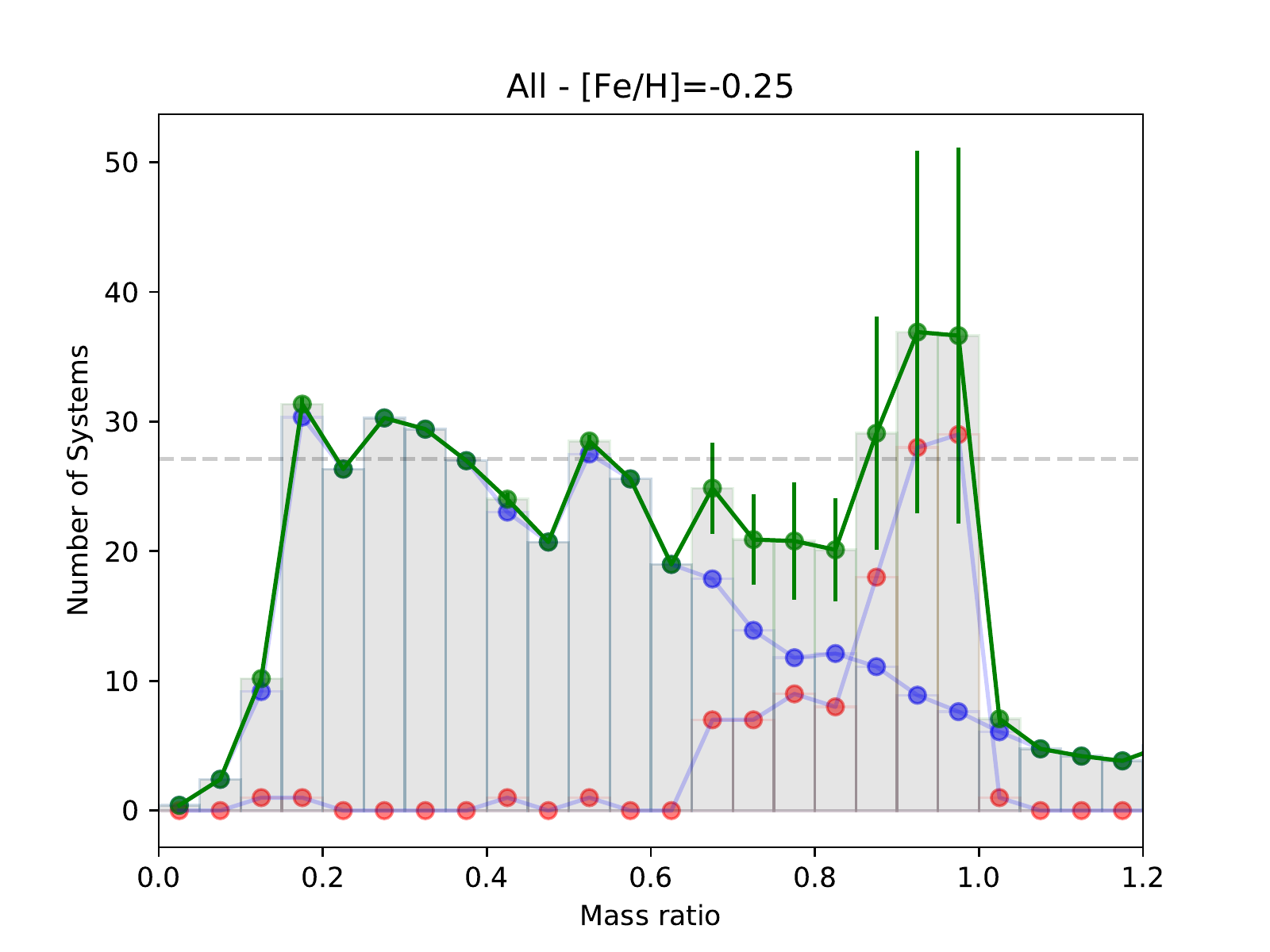}% requires the graphicx package
   \caption{The distribution of the mass ratio for all the systems we studied. SB1 and SB2 are shown separately -- as solid dots and triangles, resp., and the sum is then shown as dark stars, with the error bars indicating the range we obtain if we vary the relative number of SB2 by a factor 0.5 and 2, respectively. On top, we show the distribution when using solar-metallicity tracks, while the bottom plot shows the same when using $[Fe/H]=-0.25$ tracks.}
   \label{fig:mrall}
\end{figure}

\vspace{0.5cm}
\emph{Mass ratio distribution}
\\
For SB1s, things are more complicated, as the sole information we have is the spectroscopic mass function, $f(m)$, which is a certain combination of observables ($K_1, P, e$), and corresponds to 
\begin{equation}
f(m) = M_1 \frac{q^3}{(1+q)^2} \sin ^3 i,
\end{equation}
where $i$ is the (unknown) orbital inclination on the plane of the sky, $M_1$ is the primary mass and $q$ is the mass ratio.
If we can estimate $M_1$ or make reasonable assumptions about it, and if we can assume that $i$ is randomly distributed on the sky, that is, the probability of having a given value of $i$ is $\sin ^2 i$, then one can use statistical methods$^{7,11}$ to derive a distribution of the mass ratio -- the individual mass ratio for a given system cannot be obtained, however.

To derive $M_1$, we  made use of the \emph{Gaia} DR2 data -- our final sample consisting of those stars for which the consortium working on the data was able to derive the effective temperature and radius (Fig.~\ref{fig:hr}). Comparing their position in an H-R diagramme with the MESA stellar evolutionary tracks as obtained with MIST$^{12}$ allows us to derive the stellar mass, using a least-square method. The result is shown in Fig.~\ref{fig:mrall}, assuming all stars have solar metallicities. This figure shows nicely  how the mass ratio distribution of SB1 complements that of SB2, although there is possibly a deficit of systems with mass ratios between 0.6 and 0.8, as well as the well-known deficit of systems with $q < 0.1$, the latter being the ``brown dwarf desert". 

Assuming solar metallicities is of course a very crude approximation, which is certainly not correct -- for some stars in our sample, \emph{Simbad} gives values for the metallicity, and some are clearly sub-solar, although the average over the sample is apparently not very far from solar. Stars which are metal deficient will follow a different path in the H-R diagramme, and although the effect is negligible on the main sequence, it will lead to a shift on the red giant branches. Assuming solar metallicity will thus lead to an incorrect primary mass, especially for the red giants in our sample. It would therefore be better to use stellar evolutionary tracks of the metallicity of each star in our sample in order to derive the primary mass. Unfortunately, this information is not yet available for the majority of our sample, and we need to accept  resorting to a simplified method. As the mass ratio  depends only on the cubic root of the primary mass, we can, however, hope that the resulting effect is not very large. To test this, we have also determined the primary mass, using $[Fe/H]=-0.25$ tracks, and determined the new resulting mass ratio distribution. The results are shown in the bottom plot of Fig.~\ref{fig:mrall}, where one can see that the differences with the upper panel are minimal, with the most striking difference being that the distribution is now closer to an uniform one. 

\begin{figure}[htbp]
   \centering
   \includegraphics[width=0.8\textwidth]{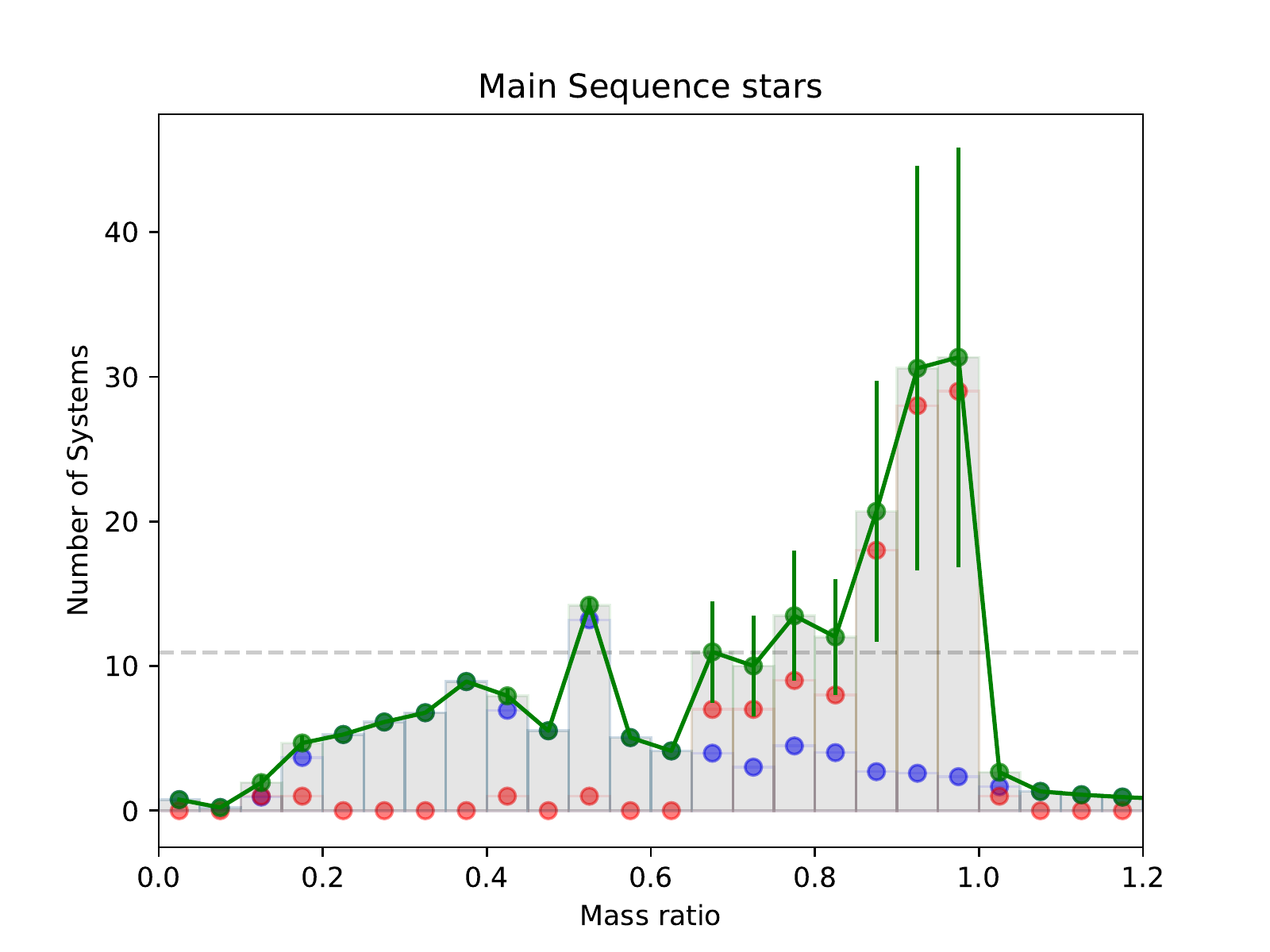}\\
   \includegraphics[width=0.8\textwidth]{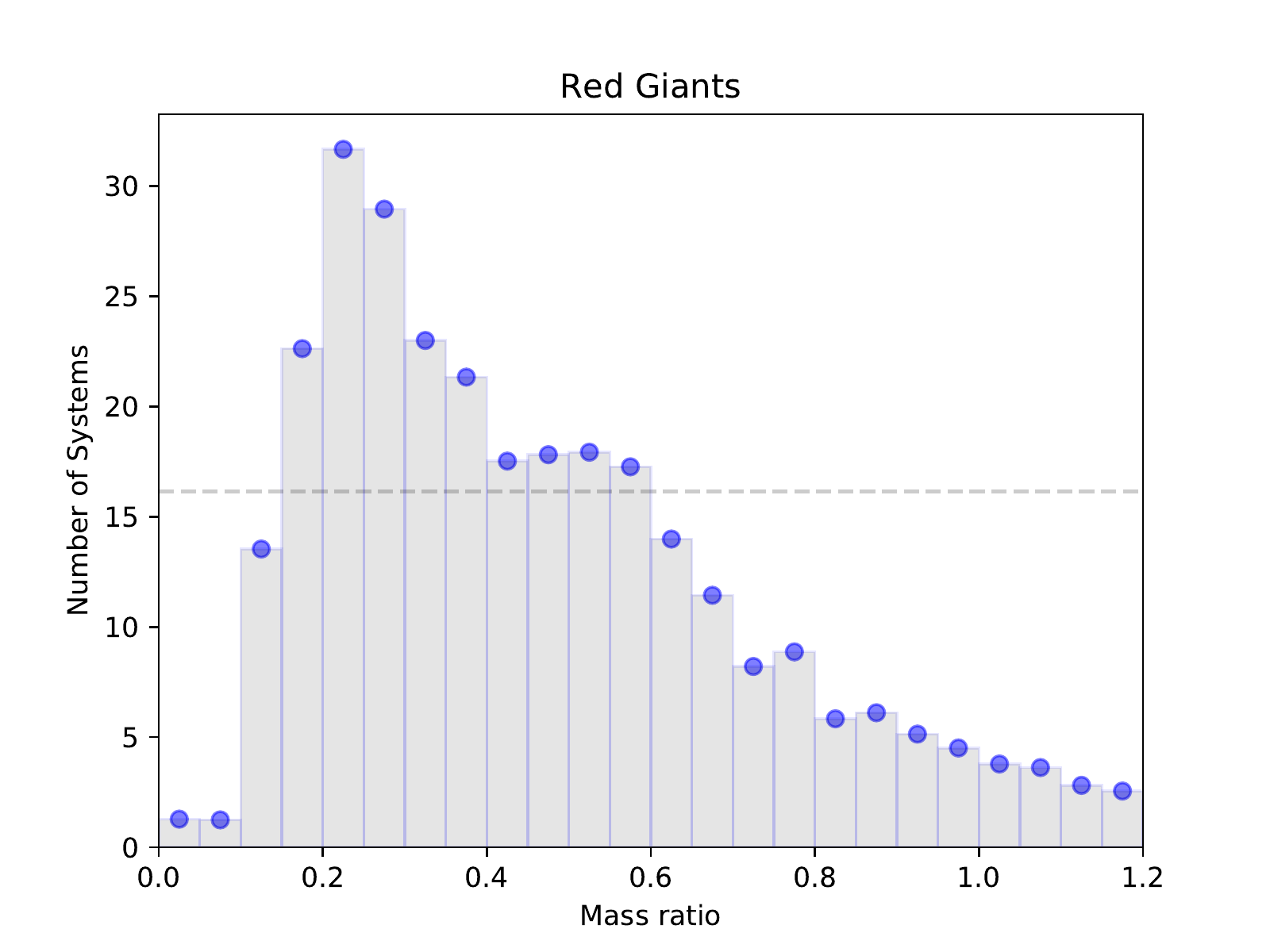}% requires the graphicx package
   \caption{Same as Fig.~\ref{fig:mrall}, separating the sample into main-sequence stars (top) and red giants (bottom).}
   \label{fig:mrsub}
\end{figure}

To obtain the final mass ratio distribution it is obviously necessary to combine the distribution of mass ratios obtained for SB1 and those for SB2 systems. The simplest way to do so is to assume that our sample is not affected by any bias that would alter the ratio between SB2 and SB1 systems. Given the methodology of Griffin, this seems to be a very reasonable assumption, as the selection is mostly done on the fact that the system has a spectral type that matches the \emph{Coravel's} mask and is bright enough to be observed. A possible bias would be that an SB2 may appear on average brighter than an SB1 as the companion contributes light. However, when looking at the distribution of magnitudes, it appears that SB2 from the current sample are, on average, fainter than SB1s. This bias is thus not operating. 
Nevertheless, in order to show what would be the effect of any bias affecting the ratio between SB1 and SB2 systems, we indicate with error bars in Fig.~\ref{fig:mrall} the effect of multiplying or dividing by two the contribution of SB2 systems to the final distribution.

\vspace{0.5cm}
\emph{Looking in more detail}
\\
In order to see if the features we detect in our final mass-ratio distribution are due to a particular population, we have split our final sample into main-sequence stars and red giants and determined the mass ratio distributions for these subsamples -- this is done in Fig.~\ref{fig:mrsub}.
When doing so, it is now very clear that the peak of systems with mass ratios above 0.85 is restricted to main-sequence stars only. As mentioned before, in fact this is mostly due to a population of F-type stars in our sample. For main-sequence stars, we also clearly see the brown dwarf desert at the lowest mass ratios. It is important to note that given the smallest semi-amplitudes that Griffin could detect, such brown dwarf can be detected (in fact, some have), so their paucity must be a reality. There is also an apparent peak of systems with mass ratio around 0.5, but its significance is barely at the 3-sigma level and it is probably better to not give it too much attention.

Turning now to the mass-ratio distribution of red giants, it is clear that there is a decline of systems with a mass ratio larger than about 0.6. This result was also found by Van der Swaelmen et al.$^{13}$ in a less pronounced way when studying a sample of red giant members of open clusters. Its origin must lie in the fact that the red giants in our sample originate mostly from stars with masses around 2-2.5 M$_\odot$, i.e. evolved A-type stars, unlike our sample of main-sequence stars which is comprised of mostly G and F-type stars. This could thus point to the fact that A-type stars have an inherent deficit of relatively massive companions.  

The mass-ratio distribution of red giants also shows a clear peak around a mass ratio 0.25--0.3, which was again seen by Van der Swaelmen et al.$^{13}$. They conclude that 22\% of their sample corresponds to post-mass-transfer systems, in which the red giant's companion is a white dwarf. This result is in agreement with what we find here as well.

%\pagebreak
\vspace{0.5cm}
\emph{Acknowledgements}
\\
It is an immense pleasure to thank Prof. R. F. Griffin for having collected patiently so many orbits in his \emph{Series} in this \emph{Magazine}, and for allowing us to perform the current analysis of his systems.
This research has made use of the SIMBAD database,
operated at CDS, Strasbourg, France and of the VizieR catalogue access tool, CDS, Strasbourg, France (DOI: 10.26093/cds/vizier). The original description of the VizieR service was published in A\&AS 143, 23. 
This work has made use of data from the European Space Agency (ESA) mission
{\it Gaia} \\(\url{https://www.cosmos.esa.int/gaia}), processed by the {\it Gaia}
Data Processing and Analysis Consortium (DPAC,\\
\url{https://www.cosmos.esa.int/web/gaia/dpac/consortium}). Funding for the DPAC
has been provided by national institutions, in particular the institutions
participating in the {\it Gaia} Multilateral Agreement.

\vspace{0.5cm}
\begin{center}
{\it References}
\end{center}

\setlength{\parindent}{0pt}
\begin{itemize}
\item[]  (1) R. F. Griffin, A Photometric Atlas of the Spectrum of Arcturus, {\it Cambridge Philosophical Society}, Cambridge, 1968.
\item[]  (2) R. Griffin, R. Griffin, A photometric atlas of the spectrum of Procyon $\lambda \lambda 3140-7470$, A.Cambridge: Institute of Astronomy, 1979.
\item[]  (3) R. Griffin, {\it MNRAS}, 162, 243, 1973.
\item[]  (4) R. F. Griffin, {\it ApJ}, 148, 465, 1967.
\item[]  (5) R. F. Griffin, B. Emerson, {\it The Observatory}, 95, 23, 1975.
\item[]  (6) V. Trimble, {\it The Observatory}, 128, 286, 2008.
\item[]  (7) H. M. J. Boffin, N. Cerf, G. Paulus, {\it A\& A}, 271, 125, 1993. 
\item[]  (8) R. F. Griffin, {\it The Observatory}, 128, 448, 2008.
\item[]  (9) Gaia Collaboration, Prusti, T., de Bruijne, J.~H.~J. et al., {\it A\& A}, 595, A1, 2016.
\item[] (10) Gaia Collaboration, Brown, A.~G.~A., Vallenari, A. et al., {\it A\& A}, 616, A1, 2018. 
\item[] (11) H. M. J. Boffin, {\it A\& A}, 524, A14, 2010.
\item[] (12) J. Choi et al, {\it ApJ}, 823, 102, 2016.
\item[] (13) M. Van der Swaelmen, et al., {\it A\& A}, 597, A68, 2017.
\end{itemize}

\end{document}